\documentclass[preprint,preprintnumbers,amsmath,amssymb,nofootinbib]{revtex4}
\pdfoutput=1
\usepackage{bbm}
\usepackage{amsfonts}
\usepackage{booktabs}
\usepackage{mathrsfs}
\usepackage{epsfig}
\usepackage{graphicx}
\usepackage{dcolumn}
\usepackage{bm}
\usepackage{amsmath}
\usepackage{slashed}
\usepackage{color}

\let\jnfont=\rm
\def\NPB#1,{{\jnfont Nucl.\ Phys.\ B }{\bf #1},}
\def\PLB#1,{{\jnfont Phys.\ Lett.\ B }{\bf #1},}
\def\EPJC#1,{{\jnfont Eur.\ Phys.\ Jour.\ C }{\bf #1},}
\def\PRD#1,{{\jnfont Phys.\ Rev.\ D }{\bf #1},}
\def\PRL#1,{{\jnfont Phys.\ Rev.\ Lett.\ }{\bf #1},}
\def\MPLA#1,{{\jnfont Mod.\ Phys.\ Lett.\ A }{\bf #1},}
\def\JPG#1,{{\jnfont J.\ Phys.\ G}{\bf #1},}
\def\CTP#1,{{\jnfont Commun.\ Theor.\ Phys.\ }{\bf #1},}
\def\ZPC#1,{{\jnfont Z.\ Phys.\ C }{\bf #1},}
\def\JHEP#1,{{\jnfont JHEP \ }{\bf #1},}
\def\Rv{\not{\hbox{\kern-1pt $R$}}}
\def\p{\not{\hbox{\kern-3pt $p$}}}

\begin{document}

\title{Status and prospects of light bino-higgsino dark matter in natural SUSY}

\author{Murat Abdughani$^{2,3}$ 
,
Lei Wu$^{1}$ \footnote[2]{Email: leiwu@itp.ac.cn}
,
Jin Min Yang$^{2,3}$ 
}
\affiliation{
$^1$ Department of Physics and Institude of Theoretical Physics, Nanjing Normal University, Nanjing, Jiangsu 210023, China\\
$^2$ CAS Key Laboratory of Theoretical Physics, Institute of Theoretical Physics, Chinese Academy of Sciences, Beijing 100190, China\\
$^3$ School of Physical Sciences, University of Chinese Academy of Sciences, Beingjing 100049, China}

\small
\begin{abstract}
Given the recent progress in dark matter direction detection experiments, we examine a light bino-higgsino dark matter (DM) scenario ($M_1<100$ GeV and $\mu<300$ GeV) in natural supersymmetry with the electroweak fine tuning measure $\Delta_{EW}<30$. By imposing various constraints, we note that: (i) For $sign(\mu/M_1)=+1$, the parameter space allowed by the DM relic density and collider bounds can almost be excluded by the very recent spin-independent (SI) scattering cross section limits from the XENON1T (2017) experiment. (ii) For $sign(\mu/M_1)=-1$, the SI limits can be evaded due to the cancelation effects in the $h\tilde{\chi}^0_1\tilde{\chi}^0_1$ coupling, while rather stringent constraints come from the PandaX-II (2016) spin-dependent (SD) scattering cross section limits, which can exclude the higgsino mass $|\mu|$ and the LSP mass $m_{\tilde{\chi}^0_1}$ up to about 230 GeV and 37 GeV, respectively. Furthermore, the surviving parameter space will be fully covered by the projected XENON1T experiment or the future trilepton searches at the HL-LHC.

\end{abstract}
\maketitle

\normalsize
\section{INTRODUCTION}
Scrutinizing the mechanism for stabilizing the electroweak scale becomes more impending after the Higgs discovery at the LHC~\cite{higgs-atlas,higgs-cms}. Besides, there is overwhelming evidence for the existence of dark matter from cosmological observations. Identifying the nature of dark matter is one of the challenges in particle physics and cosmology.

The weak scale supersymmetry is widely regarded as one of the most appealing new physics models at the TeV scale. It can successfully solve the naturalness problem in the Standard Model (SM) and also provide a compelling cold dark matter candidate. Among various supersymmetric models, the natural supersymmetry is a well motivated framework (see examples \cite{nsusy-1,nsusy-2,nsusy-3,nsusy-4,nsusy-5,nsusy-7,nsusy-8,nsusy-9,nsusy-10}), which usually indicates the light higgsinos in the spectrum \cite{bg}. If unification of gaugino mass parameters is further assumed, the current LHC bound on the gluino ($m_{\tilde{g}} \gtrsim 2$ TeV \cite{run2-gluino}) would imply correspondingly heavy winos and binos, resulting in a higgsino-like lightest supersymmetric particle (LSP). However, the thermal abundance of light higgsino-like LSP is typically lower than the observed value of the dark matter in the universe, due to the large higgsino-higgsino annihilation rate. These considerations motivate us to explore the phenomenology of neutralino dark matter in natural SUSY by giving up the gaugino mass unification assumption. One of the possibilities is to allow for the light bino in natural SUSY. Such a mixed bino-higgsino neutralino dark matter can solve the above mentioned problems of a pure higgsino LSP without worsening the naturalness in natural SUSY. The studies of bino-higgsino dark matter have also been carried out in  \cite{bh-1,bh-2,bh-3,bh-4,bh-5,bh-6,bh-7,bh-8,bh-9,bh-10,bh-11,bh-12,bh-13,bh-14,bh-15,bh-16,bh-17,bh-18,gambit-1,gambit-2}.

In this work, we will confront the light bino-higgsino dark matter scenario in natural SUSY with the recent direct detection data. In particular, we focus on the light dark matter regime ($m_{\tilde{\chi}^0_1}<100$ GeV) and attempt to address the lower limit of the mass of LSP that saturates the dark matter relic abundance. In natural SUSY, a small $\mu$ parameter leads to a certain bino-higgsino mixing, so that the spin-independent/dependent neutralino LSP-nucleon scattering cross sections can be enhanced. We will utilize the recent XENON1T \cite{xenon1t} and PandaX-II \cite{pandax} limits to examine our parameter space. Since the couplings of the LSP with the SM particles depend on the relative sign ($sign(\mu/M_1)$) between the mass parameters $\mu$ and $M_1$, we will include both of $sign(\mu/M_1)=\pm 1$ in our study and show its impact on the exclusion limits for our scenario. Besides, we explore the potential to probe such a scenario by searching for the trilepton events at 14 TeV LHC.

The structure of this paper is organized as follows. In Section \ref{section2}, we will discuss the light bino-higgsino neutralino parameter space in natural SUSY. In Section \ref{section3}, we will perform the parameter scan and discuss our numerical results. Finally, we draw our conclusions in Section \ref{section4}.

\section{light bino-higgisino neutralino in natural SUSY}\label{section2}
In the MSSM, the minimization of the tree-level Higgs potential leads to the following equation \cite{mz}
\begin{eqnarray}
\frac{M^2_{Z}}{2}&=&\frac{m^2_{H_d}-m^2_{H_u}\tan^{2}\beta}{\tan^{2}\beta-1}-\mu^{2},
\label{minimization}
\end{eqnarray}
where $m^2_{H_{u,d}}$ denote the soft SUSY breaking masses of the Higgs fields at the weak scale, respectively. It should be noted that the radiative EWSB condition usually imposes a non-trivial relation between the relevant soft mass parameters at the high scale in a UV model, such as mSUGRA. However, the scenario we studied in our work is the so-called low energy phenomenological MSSM, in which a successful EWSB is always assumed and in this case the above mentioned strong correlation between parameters from radiative EWSB condition in UV models is not applicable. Using the electroweak fine tuning measure $\Delta_{EW}$~\cite{nsusy-4}, one can see that the higgsino mass parameter $\mu$ should be of the order of $\lesssim 300$ GeV to satisfy the requirement of $\Delta_{EW} < 30$~\cite{ft-1,ft-2,ft-3,higgsino-1}. The light higgsinos have been searched for through chargino pair production in the LEP-2 experiment \cite{lep2}, which indicates $\mu \gtrsim 100$ GeV. We will use this LEP-2 limit as a lower bound for the higgsino mass. However, the relic abundance of thermally produced pure higgsino LSP falls well below dark matter measurements, unless its mass is in the TeV range. In order to provide the required relic density, several alternative ways have been proposed, such as the multi-component dark matter that introducing the axion \cite{axion}. On the other hand, without fully saturating the relic density (under-abundance), the higgsino-like neutralino dark matter in radiatively-driven natural supersymmetry with $\Delta_{EW} < 30$ \cite{hwimp} or natural mini-landscape \cite{minilandscape} has been confronted with various (in-)direct detections and is also expected to be accessible via Xenon1T experiment. In our study, we achieve the correct dark matter relic density by allowing the light bino to mix with the higgsinos.

The two neutral higgsinos ($\tilde{H}_u^0$ and $\tilde{H}_d^0$) and the two neutral guaginos ($\tilde{B}$ and $\tilde{W}^0$) are combined to form four mass eigenstates called neutralinos. In the gauge-eigenstate basis ($\tilde{B}$, $\tilde{W}^0$, $\tilde{H}_d$, $\tilde{H}_u$), the neutralino mass matrix takes the form:
\begin{equation}
M_{\tilde{\chi}^0} = \left(
\begin{matrix}
       M_1         &     0            & -c_\beta s_W m_Z & s_\beta s_W m_Z  \\
       0           &     M_2          & c_\beta c_W m_Z  & -s_\beta s_W m_Z \\
  -c_\beta s_W m_Z &  c_\beta c_W m_Z &    0             &     -\mu         \\
  s_\beta s_W m_Z  & -s_\beta s_W m_Z &    -\mu          &      0
\end{matrix}
\right)
\label{eq:neutralinomatrix}
\end{equation}
where $s_\beta = \sin\beta$, $c_\beta = \cos\beta$, $s_W = \sin\theta_W$, $c_W = \cos\theta_W$,
$M_1$ and $M_2$ are the soft-breaking mass parameters for bino and wino, respectively.
$M_{\tilde{\chi}^0}$ can be diagonalized by a $4\times 4$ unitary matrix $N$. In the limit of $M_1<\mu \ll M_2$,
the lightest neutralino is bino-like (with some higgsino mixture),
while the second and third neutralinos are higgsino-like.
The LSP can interact with nuclei via exchange of squarks and Higgs bosons (spin-independent) and
via exchange of $Z$ boson and squarks (spin-dependent). Given the strong LHC bounds on the squarks and non-SM
Higgs bosons, one can neglect their contributions to the scattering cross section. Then, the couplings of the LSP
with the Higgs boson can be written by
\begin{eqnarray}
C_{h\tilde{\chi}^0_1\tilde{\chi}^0_1}& \approx & -\sqrt{2}g_1 N_{11}^2 \frac{M_Z s_W}{\mu}
\,\frac{M_1/\mu + \sin2\beta}{1-\left(M_1/\mu\right)^2}.
\label{coupling}
\end{eqnarray}
where $N_{11}$ denotes the bino component of the lightest neutralino mass eigenstate. It can be seen that the SI scattering cross section depends on the relative sign of $M_1$ and $\mu$.
When $sign(M_1/\mu)<0$, the coupling $C_{h\tilde{\chi}^0_1\tilde{\chi}^0_1}$ can be suppressed and even vanish
if $M_1/\mu =-\sin2\beta$ so that the strong LUX SI limits can be escaped. For the SD scattering cross section,
it should be noted that the coupling $Z\tilde{\chi}^0_1\tilde{\chi}^0_1$ can appear via the higgsino component
in the LSP. The pure bino/wino LSP will not have interactions with $Z$ boson, while the pure higgsino LSP can
only have the non-zero coupling $Z\tilde{\chi}^0_1\tilde{\chi}^0_2$. Another blind spot in SD scattering
can happen in the limit of $\tan\beta=1$, where the left-right parity is restored and the parity-violating $Z$
coupling will vanish \cite{bh-3}. However, a low value of $\tan\beta$ is disfavored by the observed Higgs mass
in the MSSM.

\section{Parameter Scan and Numerical results}\label{section3}
In our numerical calculations, we vary the relevant parameters in the ranges of
\begin{eqnarray}
100~{\rm GeV} \le |\mu| \le 300~{\rm GeV},\  30~{\rm GeV}  \le |M_1| \le 100~{\rm GeV},\  10 \le tan\beta \le 50.
\end{eqnarray}
We scan the values of $M_1$ up to 100 GeV since we are interested in light DM region and attempt to address the lower limit of the LSP mass. For higher upper values of $\mu$ and $M_1$, a heavy mixed higgsino-bino LSP may also produce the right DM relic abundance~\cite{bh-8}, while the result for lower bound of LSP mass obtained in the following calculation will not change. The stop and gluino can contribute to the naturalness at loop level, which are expected to be $m_{\tilde{t}_1} \lesssim 2.5$ TeV and $m_{\tilde{g}} \lesssim 3-4$ TeV for $\Delta_{EW} <30$ \cite{ft-1,th-bound-1}. By recasting the LHC Run-2 with $\sim 15$ fb $^{-1}$ of data, it is found that the lower bounds of stop mass and gluino mass are about 800 GeV \cite{nsusy-stop,nsusy-stop-0,nsusy-stop-1,nsusy-stop-2,nsusy-stop-3,nsusy-stop-4} and 1.5 TeV  \cite{nsusy-stop-gluino} in natural SUSY, respectively. Given the irrelevance of the third generation parameters for our neutralino dark matter, we fix the third generation squark soft masses as $M_{\tilde{Q}_{3L}}=3$ TeV, $M_{\tilde{t}_{3R}}=M_{\tilde{b}_{3R}}=1$ TeV and vary the stop trilinear parameters in the range $|A_{t}|<2$ TeV for simplicity. The physical stop mass $m_{\tilde{t}_1}$ has to be less than 2.5 TeV to satisfy $\Delta_{EW} <30$. We also require that each sample can guarantee the correct Higgs mass and the vacuum stability \cite{stability-1,stability-2}. The first two generation squark and all slepton soft masses are assumed to be 3 TeV. Other trilinear parameters are fixed as $A_{f}=0$. We also decouple the wino and gluino by setting $M_{2,3} = 2$ TeV. We impose the following constraints in our scan:
\begin{itemize}

\item[(1)] The light CP-even Higgs boson masses of our samples should be within the range of 122--128 GeV. The package \textsf{SuSpect} \cite{suspect} is used to calculate the Higgs mass.

\item[(2)] The samples have to be consistent with the Higgs data from LEP, Tevatron and LHC. We use the package \textsf{HiggsBounds-4.2.1} \cite{higgsbounds} and \textsf{HiggsSignals-1.4.0} \cite{higgssignals} to implement the constraints.

\item[(3)] The relic density of neutralino dark matter $\Omega_{\tilde{\chi}}h^2$ is computed by \textsf{MicrOMEGAs 4.3.2} \cite{micromega}. Including 10\% theoretical uncertainty, we require our samples to satisfy the observed value $0.1186\pm 0.0020$ \cite{micromega} within $2\sigma$ range.

\item[(4)] If $m_{\tilde{\chi}^0_1}<m_h/2$, the SM Higgs boson can decay to $\tilde{\chi}^0_1\tilde{\chi}^0_1$ invisibly. We require the branching ratio $Br(h \to \tilde{\chi}^0_1\tilde{\chi}^0_1)<24\%$, which has been recently given by CMS collaboration at 95\% C.L. \cite{cms-invisible}.

\item[(5)] The invisible width of the $Z$ boson is required less than 0.5 MeV to satisfy the LEP limit.

\item[(6)] The LEP searches for $\tilde{\chi}^0_1\tilde{\chi}^0_{2,3}$ associated production gives an upper limit, $\sigma(e^+e^- \to \tilde{\chi}^0_1\tilde{\chi}^0_{2,3} \times Br(\tilde{\chi}^0_{2,3}\to \tilde{\chi}^0_{1}Z^*)<100$ fb.
\end{itemize}

\begin{figure}[h]
\centering
\includegraphics[width=6in,height=3.5in]{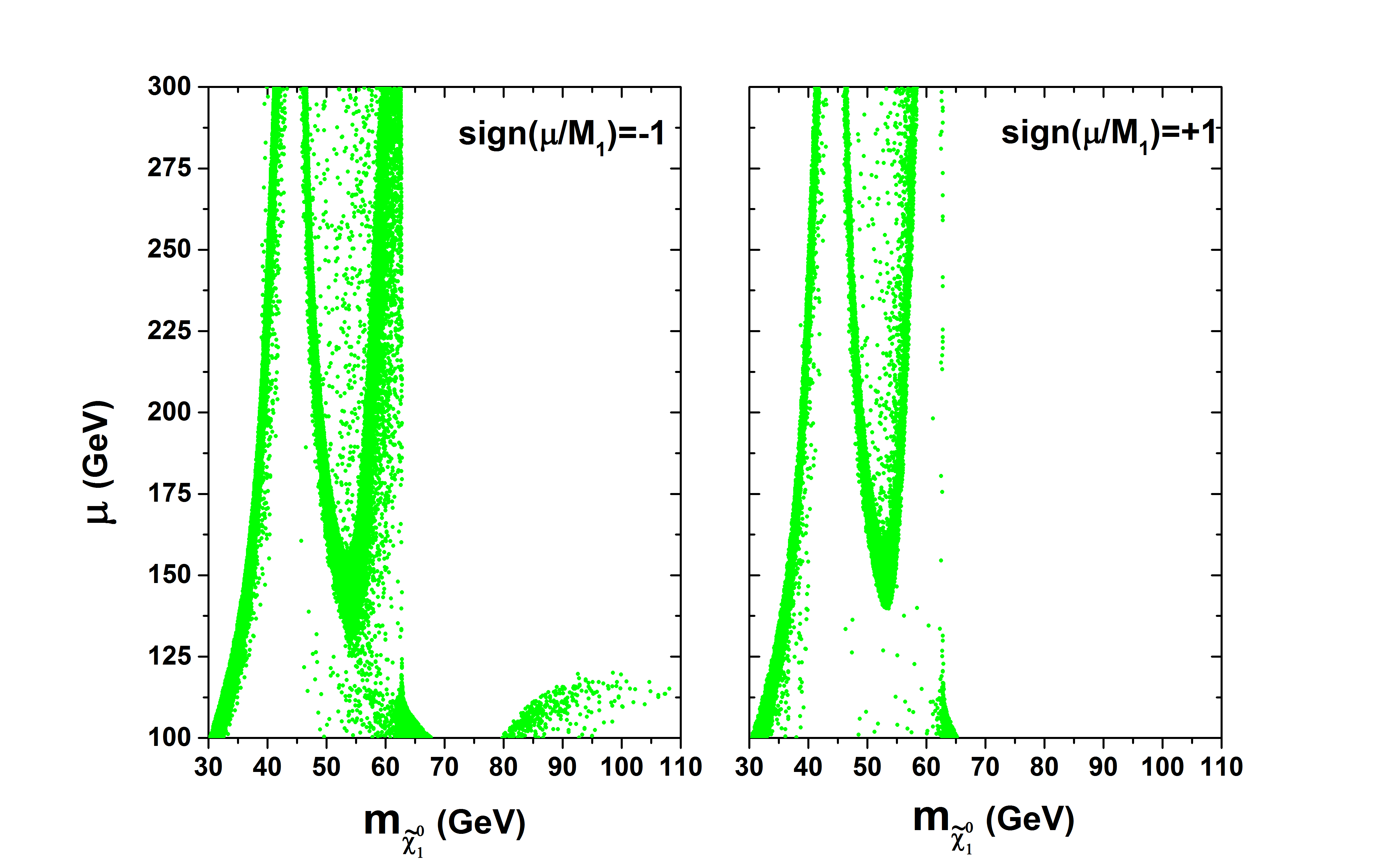}
\vspace{-0.5cm}
\caption{Scattering plot of samples satisfying the dark matter relic density.}
\label{fig:omega}
\end{figure}
In Fig.~\ref{fig:omega}, we show the samples satisfying the dark matter relic density for $sign(\mu)=\pm1$.
Since a bino-like LSP  has rather small couplings with the SM particles,
a certain portion of higgsino components is required to meet the observed relic density.
Otherwise, the universe will be overclosed. Therefore, except for the two resonance regions
$m_{\tilde{\chi}^0_1} \simeq m_Z/2$ and $m_h/2$, the higgsino mass parameter $\mu$ is expected to be as
low as possible in our scan ranges. It should be noted that the difference of $sign(\mu/M_1)=\pm 1$
in calculating the relic abundance mainly happens around and after the Higgs resonance region, in which
more samples are allowed for $sign(\mu/M_1)=-1$. This is because that the negative sign of $\mu/M_1$ can
reduce the coupling of the LSP with the Higgs boson and the suppress the enhanced annihilation cross section
of $\tilde{\chi}^0_1\tilde{\chi}^0_1$ by the Higgs resonant effect. When $m_{\tilde{\chi}^0_1} > m_h/2$, the LSP
for $sign(\mu/M_1)=\pm 1$ is still bino-like so that the relic density easily exceeds the observed value.
But if $M_1$ is close to $\mu$, the LSP for $sign(\mu/M_1)=-1$ can have sizable higgsino components, which
allows samples in the lower right corner on the left panel of Fig.~\ref{fig:omega}. However, such a region
will be excluded by the dark matter direct detections as shown in the following.

\begin{figure}[h]
\centering
\includegraphics[width=6in,height=6in]{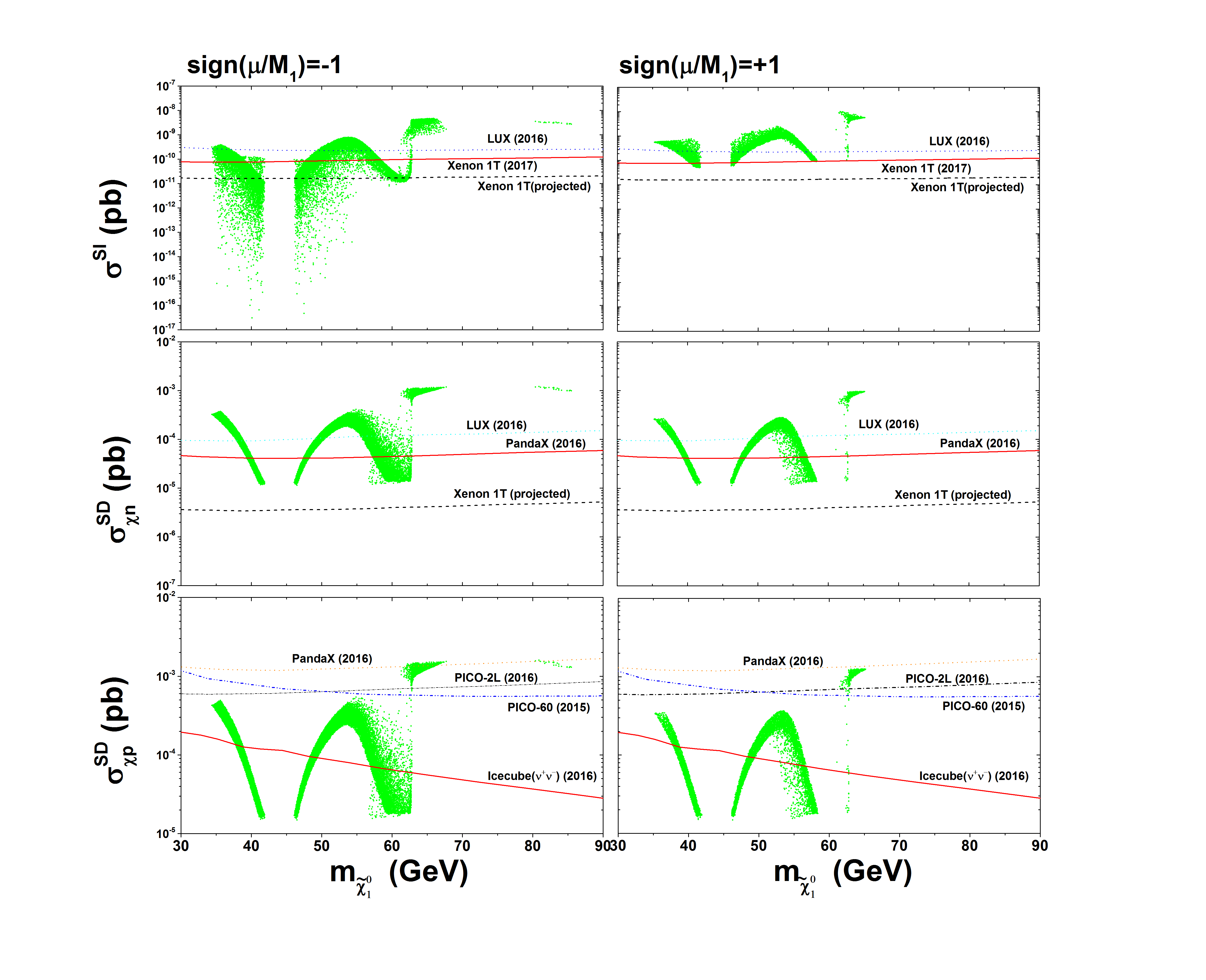}
\vspace{-1.5cm}
\caption{Spin-independent/dependent neutralino LSP-nucleon scattering cross sections. All samples satisfying the
constraints (1-6). The observed 90\% C.L. upper limits from Xenon1T (2017) \cite{xenon1t}, PandaX (2016) \cite{pandax},
LUX (2016) \cite{lux_sd,lux_si}, PICO-2L (2016) \cite{pico2l}, PICO-60 (2015) \cite{pico60}, IceCube (2016) \cite{icecube}
and the projected XENON1T sensitivity limits \cite{pxenon1t} are plotted. For indirect limits, we assume that LSP
annihilates exclusively to some specific final state, with a canonical thermal annihilation cross-section
$\langle\sigma v\rangle_0=3\times 10^{-26} cm^3 s^{-1}$.}
\label{fig:DD}
\end{figure}
In Fig.~\ref{fig:DD}, we present the spin-independent/dependent neutralino LSP-nucleon scattering cross sections,
which are calculated by using \textsf{MicrOMEGAs 4.3.2} \cite{micromega}. All samples satisfying the constraints (1-6).
The neutron and proton form factors are taken as $f^p_d \approx 0.132$ and $f^n_d \approx 0.140$. It can be seen that
the very recent SI cross section limits from XENON1T experiment can almost exclude the whole parameter space
of $sign(\mu/M_1)=+1$. While for $sign(\mu/M_1)=-1$, a large portion of our samples can escape the SI limits
since the $h\tilde{\chi}^0_1\tilde{\chi}^0_1$ coupling is suppressed by the cancelation effect in Eq.~(\ref{coupling}).

On the other hand, the SD cross section is largely determined by $Z$-boson exchange and is sensitive to the higgsino
asymmetry, $\sigma_{SD} \propto |N^{2}_{13}-N^{2}_{14}|^2$. The relic density constraint requires a large higgsino asymmetry
so that the SD cross section is enhanced. Therefore, a strong bound on such a scenario comes from the PandaX-II (2016)
SD neutralino LSP-neutron scattering cross section limits, which can rule out about 70\% of our samples and exclude the higgsino mass $|\mu|$ and the LSP mass $m_{\tilde{\chi}^0_1}$ up to about 230 GeV and 37 GeV, respectively. Such lower limits will not changed even if we extend the scan ranges of $M_1$ and $\mu$ to larger values. The current SD neutralino LSP-proton limits from PandaX and
PICO are still weak. Both of $sign(\mu)=\pm 1$ scenarios can be completely covered by the projected XENON1T experiment
in the future.

Besides the direct detections, the neutralino annihilation in the Sun to neutrinos can also be enhanced by the
higgsino component in the LSP. The null results from the neutrino telescopes, such as IceCube, have produced a strong
bound on the SD neutralino LSP-proton scattering cross sections and has excluded a sizable portion of the parameter
space for $sign(\mu)=-1$. Next, we discuss the LHC potential of probing the current parameter space of our scenario allowed by the constraints (1-6)
and the above direct/indirect detections.

\begin{figure}[ht]
\centering
\includegraphics[width=5in,height=3.5in]{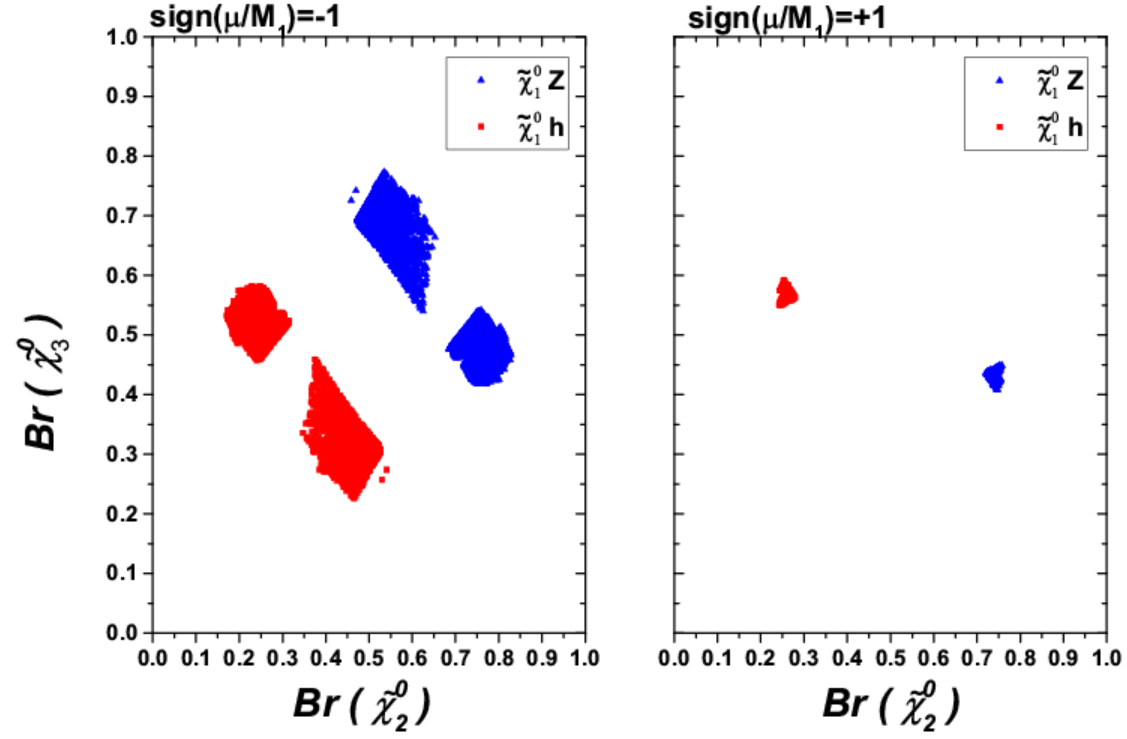}
\vspace{-0.5cm}
\caption{Scatter plots of the samples
allowed by the constraints (1-6) and by the XENON1T (2017) and PandaX (2016),
showing $\tilde\chi_2^0$ and $\tilde\chi_3^0$ decay branching ratios.}
\label{fig:bratio}
\end{figure}
In Fig.~\ref{fig:bratio}, we plot the decay branching ratios of $\tilde\chi_2^0$
and $\tilde\chi_3^0$. For $sign(\mu)=-1$, we can see that the neutralinos $\tilde{\chi}^0_{2,3}$ mainly decay
to $\tilde{\chi}^0_1 Z$. When $Br(\tilde{\chi}^0_2 \to \tilde{\chi}^0_1 Z)$ increases,
$Br(\tilde{\chi}^0_3 \to \tilde{\chi}^0_1 Z)$ decreases because of the goldstone theorem \cite{bh-12}.
A similar correlation can be seen in the decay channel $\tilde{\chi}^0_{2,3} \to \tilde{\chi}^0_1 h$.
But for $sign(\mu)=+1$, the neutralino $\tilde{\chi}^0_{2}$ still dominantly decay to $\tilde{\chi}^0_1 Z$,
while the neutralino $\tilde{\chi}^0_{3}$ preferently decay to $\tilde{\chi}^0_1 h$. This indicates that the samples with negative sign of $\mu/M_1$ will produce more trilepton events through the process $pp \to \tilde{\chi}^0_{2,3}(\to Z\tilde{\chi}^0_1)\tilde{\chi}^\pm_1(\to W^\pm \tilde{\chi}^0_1)$ than those with positive sign of $\mu/M_1$, and can be more easily excluded by the null results of searching for electroweakinos at the LHC.

\begin{table}[th]
\caption{Recasted LHC-8 TeV analyses with 20.3~fb$^{-1}$ of data and corresponding signals in our scenario.}
 \vspace*{0.5cm}
\begin{tabular}{|c|c|}
\hline
Final states & Source of signal in our scenario\\
\hline
$3 {\rm lepton} + \slashed E_T$ \cite{atlas-1402-7029}  & $pp \to \tilde{\chi}^\pm_1(\to W^\pm \tilde{\chi}^0_1)\tilde{\chi}^0_{2,3}(\to Z \tilde{\chi}^0_1)$ \\
\hline
$1 {\rm lepton} + h + \slashed E_T$ \cite{atlas-1501-07110} & $pp \to \tilde{\chi}^\pm_1(\to W^\pm \tilde{\chi}^0_1)\tilde{\chi}^0_{2,3}(\to h \tilde{\chi}^0_1)$ \\
\hline
$\ell^+\ell^- + \slashed E_T$ \cite{atlas-conf-2013-049} & $pp \to \tilde{\chi}^+_1\tilde{\chi}^-_1$\\
\hline
\end{tabular}
\label{run1}
\end{table}

Given the above decay modes, we first recast the LHC searches for the electroweakinos listed in Table ~\ref{run1}
with \textsf{CheckMATE2} \cite{checkmate}.
We generate the parton level signal events by \textsf{MadGraph5\_aMC@NLO} \cite{mad5}
and perform the shower and hadronization procedure by \textsf{Pythia-8.2} \cite{pythia}. The fast detector simulation
are carried out with the tuned \textsf{Delphes} \cite{delphes}. We implement the jet clustering
by \textsf{FastJet} \cite{fastjet} with the anti-$k_t$ algorithm \cite{anti-kt}. We use \textsf{Prospino2} \cite{prospino}
to calculate the QCD corrected cross sections of the electroweakino pair productions at the LHC. Then, we estimate the
exclusion limit by evaluating the ratio $r = max(N_{S,i}/S^{95\%}_{obs,i})$, where $N_{S,i}$ is the event number of signal
for $i$-th signal region and $S^{95\%}_{obs,i}$ is the corresponding 95\% C.L. observed upper limit. A sample is excluded
at 95\% C.L. if $r > 1$. After checking all surviving samples, we find that the LHC data in Tab.~\ref{run1} can not
further exclude the parameter space because of the strong direct detection bound on higgsino mass parameter $\mu>230$ GeV.

\begin{figure}[ht]
\centering
\includegraphics[width=5in,height=3.5in]{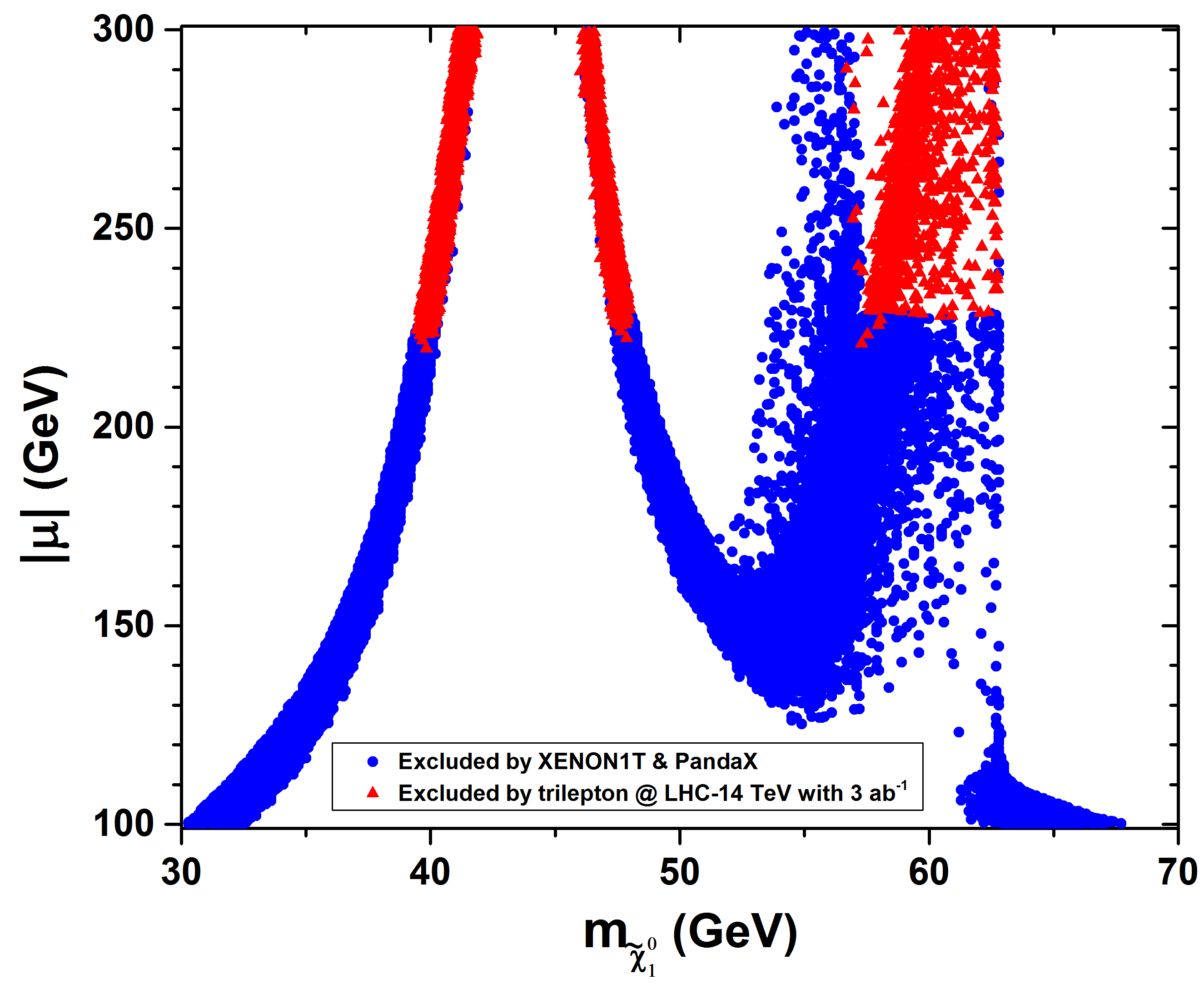}
\vspace{-0.5cm}
\caption{Scatter plots of the samples allowed by the constraints (1-6) on the plane of $|\mu|$ versus $m_{\tilde{\chi}^0_1}$.
The blue bullets are excluded by the XENON1T (2017) and PandaX (2016). The red triangle are expected to be excluded by the trilepton searches at 95\% C.L. at the HL-LHC}
\label{fig:14tev}
\end{figure}
In Fig.~\ref{fig:14tev}, we show the prospect of testing our surviving samples through searching for electroweakino
pair production in the trilepton final states at 14 TeV LHC with the luminosity ${\cal L}=3000$ fb$^{-1}$. Such an
analysis \cite{atlas-phys-2014-010-hl-3l} has been implemented in \textsf{CheckMATE} package. In order to reduce the Monte Carlo fluctuations, we generate 200,000 events for each signal point. In Fig.~\ref{fig:14tev}, we can see that all red triangles allowed by the constraints (1)-(6) and the XENON1T (2017) and PandaX (2016) experiments can be excluded by the HL-LHC at 95\% C.L.. Therefore, we conclude that our light bino-higgsino neutralino dark matter scenario will be fully tested by either future XENON1T or HL-LHC experiments.

\section{conclusion}\label{section4}
In this work, we examined light bino-higgsino neutralino dark matter in natural SUSY by imposing various constraints from the LEP, dark matter and LHC experiments. We found that the relative sign between the mass parameters
$\mu$ and $M_1$ can significantly affect the dark matter and LHC phenomenology of our scenario.
For $sign(\mu/M_1)=+1$, the very recent SI limits from the Xenon1T (2017) experiment can almost exclude
the whole parameter space allowed by the relic density and collider bounds. But for $sign(\mu/M_1)=-1$,
the SI limits can be avoided due to the cancelation effects in $h\tilde{\chi}^0_1\tilde{\chi}^0_1$ coupling.
In this case, a strong bound comes from the PandaX-II (2016) SD neutralino LSP-neutron scattering cross section limits,
which can exclude the higgsino mass $|\mu|$ and the LSP mass $m_{\tilde{\chi}^0_1}$ up to about 230 GeV and 37 GeV, respectively. Furthermore, the surviving parameter space will be fully covered by the projected XENON1T experiment or the future trilepton searches at 14 TeV LHC with the luminosity ${\cal L}=3000$ fb$^{-1}$.

\section*{Acknowledgement}
We thank G. H. Duan and Yang Zhang for helpful discussions. This work is supported by the National Natural Science
Foundation of China (NNSFC) under grant No. 11705093 and No. 11675242, by the CAS Center for Excellence in Particle Physics (CCEPP) and by the CAS Key Research Program of Frontier Sciences.


\begin{thebibliography}{99}


\bibitem{higgs-atlas}
  G.~Aad {\it et al.} [ATLAS Collaboration],
  Phys.\ Lett.\ B {\bf 716}, 1 (2012)
  doi:10.1016/j.physletb.2012.08.020
  [arXiv:1207.7214 [hep-ex]].

\bibitem{higgs-cms}
  S.~Chatrchyan {\it et al.} [CMS Collaboration],
  Phys.\ Lett.\ B {\bf 716}, 30 (2012)
  doi:10.1016/j.physletb.2012.08.021
  [arXiv:1207.7235 [hep-ex]].


\bibitem{nsusy-1}
  C.~Brust, A.~Katz, S.~Lawrence and R.~Sundrum,
  JHEP {\bf 1203}, 103 (2012).

\bibitem{nsusy-2}
  M.~Papucci, J.~T.~Ruderman and A.~Weiler,
  JHEP {\bf1209}, 035 (2012).

\bibitem{nsusy-3}
  L.~J.~Hall, D.~Pinner and J.~T.~Ruderman,
  JHEP {\bf 1204}, 131 (2012);

\bibitem{nsusy-4}
  H.~Baer, V.~Barger, P.~Huang, A.~Mustafayev and X.~Tata,
  Phys.\ Rev.\ Lett.\  {\bf 109}, 161802 (2012)
  [arXiv:1207.3343 [hep-ph]].

\bibitem{nsusy-5}
  J.~Cao, C.~Han, L.~Wu, J.~M.~Yang and Y.~Zhang,
  JHEP {\bf 1211}, 039 (2012)
  doi:10.1007/JHEP11(2012)039
  [arXiv:1206.3865 [hep-ph]].


\bibitem{nsusy-7}
  L.~Calibbi, T.~Li, A.~Mustafayev and S.~Raza,
  Phys.\ Rev.\ D {\bf 93}, no. 11, 115018 (2016)
  doi:10.1103/PhysRevD.93.115018
  [arXiv:1603.06720 [hep-ph]].


\bibitem{nsusy-8}
  F.~Wang, J.~M.~Yang and Y.~Zhang,
  JHEP {\bf 1604}, 177 (2016)
  doi:10.1007/JHEP04(2016)177
  [arXiv:1602.01699 [hep-ph]].


\bibitem{nsusy-9}
  L.~Wu,
  arXiv:1705.02534 [hep-ph].

\bibitem{nsusy-10}
  G.~H.~Duan, K.~i.~Hikasa, L.~Wu, J.~M.~Yang and M.~Zhang,
  JHEP {\bf 1703}, 091 (2017)
  doi:10.1007/JHEP03(2017)091
  [arXiv:1611.05211 [hep-ph]].


\bibitem{bg}
  R.~Barbieri and G.~F.~Giudice,
  Nucl.\ Phys.\ B {\bf 306}, 63 (1988).

\bibitem{run2-gluino}
The ATLAS collaboration, ATLAS-CONF-2015-067.



\bibitem{bh-1}
  M.~Drees and M.~M.~Nojiri,
  Phys.\ Rev.\ D {\bf 47}, 376 (1993)
  doi:10.1103/PhysRevD.47.376
  [hep-ph/9207234].

\bibitem{bh-2}
  I.~Gogoladze, R.~Khalid, Y.~Mimura and Q.~Shafi,
  Phys.\ Rev.\ D {\bf 83}, 095007 (2011)
  doi:10.1103/PhysRevD.83.095007
  [arXiv:1012.1613 [hep-ph]].


\bibitem{bh-3}
  C.~Cheung, L.~J.~Hall, D.~Pinner and J.~T.~Ruderman,
  JHEP {\bf 1305}, 100 (2013)
  doi:10.1007/JHEP05(2013)100
  [arXiv:1211.4873 [hep-ph]].


\bibitem{bh-4}
  B.~Dutta, T.~Kamon, N.~Kolev, K.~Sinha, K.~Wang and S.~Wu,
  Phys.\ Rev.\ D {\bf 87}, no. 9, 095007 (2013)
  doi:10.1103/PhysRevD.87.095007
  [arXiv:1302.3231 [hep-ph]].

\bibitem{bh-5}
  G.~B¨¦langer, G.~Drieu La Rochelle, B.~Dumont, R.~M.~Godbole, S.~Kraml and S.~Kulkarni,
  Phys.\ Lett.\ B {\bf 726}, 773 (2013)
  doi:10.1016/j.physletb.2013.09.059
  [arXiv:1308.3735 [hep-ph]].

\bibitem{bh-6}
  T.~T.~Yanagida and N.~Yokozaki,
  JHEP {\bf 1311}, 020 (2013)
  doi:10.1007/JHEP11(2013)020
  [arXiv:1308.0536 [hep-ph]].




\bibitem{bh-8}
  H.~Baer, V.~Barger, P.~Huang, D.~Mickelson, M.~Padeffke-Kirkland and X.~Tata,
  Phys.\ Rev.\ D {\bf 91}, no. 7, 075005 (2015)
  doi:10.1103/PhysRevD.91.075005
  [arXiv:1501.06357 [hep-ph]].

\bibitem{bh-7}
  C.~Han,
  arXiv:1409.7000 [hep-ph].


\bibitem{bh-9}
  A.~Kobakhidze, M.~Talia and L.~Wu,
  Phys.\ Rev.\ D {\bf 95}, no. 5, 055023 (2017)
  doi:10.1103/PhysRevD.95.055023
  [arXiv:1608.03641 [hep-ph]].

\bibitem{bh-10}
  M.~Badziak, M.~Olechowski and P.~Szczerbiak,
  Phys.\ Lett.\ B {\bf 770}, 226 (2017)
  doi:10.1016/j.physletb.2017.04.059
  [arXiv:1701.05869 [hep-ph]].

\bibitem{bh-11}
  T.~Han, F.~Kling, S.~Su and Y.~Wu,
  JHEP {\bf 1702}, 057 (2017)
  doi:10.1007/JHEP02(2017)057
  [arXiv:1612.02387 [hep-ph]].

\bibitem{bh-12}
  T.~Han, Z.~Liu and S.~Su,
  JHEP {\bf 1408}, 093 (2014)
  doi:10.1007/JHEP08(2014)093
  [arXiv:1406.1181 [hep-ph]].

\bibitem{bh-13}
  L.~Calibbi, J.~M.~Lindert, T.~Ota and Y.~Takanishi,
  JHEP {\bf 1411}, 106 (2014)
  doi:10.1007/JHEP11(2014)106
  [arXiv:1410.5730 [hep-ph]].

\bibitem{bh-14}
  J.~Kawamura and Y.~Omura,
  arXiv:1703.10379 [hep-ph].

\bibitem{bh-15}
  M.~van Beekveld, W.~Beenakker, S.~Caron, R.~Peeters and R.~Ruiz de Austri,
  arXiv:1612.06333 [hep-ph].

\bibitem{bh-16}
  M.~van Beekveld, W.~Beenakker, S.~Caron and R.~Ruiz de Austri,
  JHEP {\bf 1604}, 154 (2016)
  doi:10.1007/JHEP04(2016)154
  [arXiv:1602.00590 [hep-ph]].

\bibitem{bh-17}
  A.~Achterberg, S.~Amoroso, S.~Caron, L.~Hendriks, R.~Ruiz de Austri and C.~Weniger,
  JCAP {\bf 1508}, no. 08, 006 (2015)
  doi:10.1088/1475-7516/2015/08/006
  [arXiv:1502.05703 [hep-ph]].

\bibitem{bh-18}
  C.~Han, K.~i.~Hikasa, L.~Wu, J.~M.~Yang and Y.~Zhang,
  Phys.\ Lett.\ B {\bf 769}, 470 (2017)
  doi:10.1016/j.physletb.2017.04.026
  [arXiv:1612.02296 [hep-ph]].

\bibitem{gambit-1}
  P.~Athron {\it et al.} [GAMBIT Collaboration],
  arXiv:1705.07935 [hep-ph];

\bibitem{gambit-2}
  P.~Athron {\it et al.} [GAMBIT Collaboration],
  arXiv:1705.07917 [hep-ph];



\bibitem{xenon1t}
  E.~Aprile {\it et al.} [XENON Collaboration],
  arXiv:1705.06655 [astro-ph.CO].

\bibitem{pandax}
  C.~Fu {\it et al.} [PandaX-II Collaboration],
  Phys.\ Rev.\ Lett.\  {\bf 118}, no. 7, 071301 (2017)
  doi:10.1103/PhysRevLett.118.071301
  [arXiv:1611.06553 [hep-ex]].


\bibitem{mz}
R. Arnowitt and P. Nath, Phys. Rev. D {\bf 46}, 3981 (1992).



\bibitem{ft-1}
  H.~Baer, V.~Barger, P.~Huang, D.~Mickelson, A.~Mustafayev and X.~Tata,
  Phys.\ Rev.\ D {\bf 87}, no. 11, 115028 (2013)
  [arXiv:1212.2655 [hep-ph]].

\bibitem{ft-2}
  G.~G.~Ross, K.~Schmidt-Hoberg and F.~Staub,
  JHEP {\bf 1703}, 021 (2017)
  doi:10.1007/JHEP03(2017)021
  [arXiv:1701.03480 [hep-ph]].

\bibitem{ft-3}
  G.~G.~Ross, K.~Schmidt-Hoberg and F.~Staub,
  Phys.\ Lett.\ B {\bf 759}, 110 (2016)
  doi:10.1016/j.physletb.2016.05.053
  [arXiv:1603.09347 [hep-ph]].

\bibitem{higgsino-1}
  C.~Han, A.~Kobakhidze, N.~Liu, A.~Saavedra, L.~Wu and J.~M.~Yang,
  JHEP {\bf 1402}, 049 (2014)
  doi:10.1007/JHEP02(2014)049
  [arXiv:1310.4274 [hep-ph]].


\bibitem{lep2}
LEP2 SUSY Working Group, LEPSUSYWG/01-03.1, 2001.


\bibitem{axion}
  H.~Baer, A.~Lessa, S.~Rajagopalan and W.~Sreethawong,
  JCAP {\bf 1106}, 031 (2011)
  doi:10.1088/1475-7516/2011/06/031
  [arXiv:1103.5413 [hep-ph]].


\bibitem{hwimp}
  H.~Baer, V.~Barger and D.~Mickelson,
  Phys.\ Lett.\ B {\bf 726}, 330 (2013)
  doi:10.1016/j.physletb.2013.08.060
  [arXiv:1303.3816 [hep-ph]].


\bibitem{minilandscape}
  H.~Baer, V.~Barger, M.~Savoy, H.~Serce and X.~Tata,
  JHEP {\bf 1706}, 101 (2017)
  doi:10.1007/JHEP06(2017)101
  [arXiv:1705.01578 [hep-ph]].


\bibitem{th-bound-1}
  H.~Baer, V.~Barger, P.~Huang and X.~Tata,
  JHEP {\bf 1205}, 109 (2012).


\bibitem{nsusy-stop}
  C.~Han, K.~i.~Hikasa, L.~Wu, J.~M.~Yang and Y.~Zhang,
  JHEP {\bf 1310}, 216 (2013)
  doi:10.1007/JHEP10(2013)216
  [arXiv:1308.5307 [hep-ph]].


\bibitem{nsusy-stop-0}
  K.~Kowalska and E.~M.~Sessolo,
  Phys.\ Rev.\ D {\bf 88}, no. 7, 075001 (2013)
  [arXiv:1307.5790 [hep-ph]].


\bibitem{nsusy-stop-1}
  A.~Kobakhidze, N.~Liu, L.~Wu, J.~M.~Yang and M.~Zhang,
  Phys.\ Lett.\ B {\bf 755}, 76 (2016)
  doi:10.1016/j.physletb.2016.02.003
  [arXiv:1511.02371 [hep-ph]].

\bibitem{nsusy-stop-2}
  J.~S.~Kim, K.~Rolbiecki, R.~Ruiz, J.~Tattersall and T.~Weber,
  Phys.\ Rev.\ D {\bf 94}, no. 9, 095013 (2016)
  doi:10.1103/PhysRevD.94.095013
  [arXiv:1606.06738 [hep-ph]].


\bibitem{nsusy-stop-3}
  C.~Han, J.~Ren, L.~Wu, J.~M.~Yang and M.~Zhang,
  Eur.\ Phys.\ J.\ C {\bf 77}, no. 2, 93 (2017)
  doi:10.1140/epjc/s10052-017-4662-7
  [arXiv:1609.02361 [hep-ph]].


\bibitem{nsusy-stop-4}
  H.~Baer, V.~Barger, N.~Nagata and M.~Savoy,
  Phys.\ Rev.\ D {\bf 95}, no. 5, 055012 (2017)
  doi:10.1103/PhysRevD.95.055012
  [arXiv:1611.08511 [hep-ph]].




\bibitem{nsusy-stop-gluino}
  M.~R.~Buckley, D.~Feld, S.~Macaluso, A.~Monteux and D.~Shih,
  arXiv:1610.08059 [hep-ph].

\bibitem{stability-1}
  D.~Chowdhury, R.~M.~Godbole, K.~A.~Mohan and S.~K.~Vempati,
  JHEP {\bf 1402}, 110 (2014)
  doi:10.1007/JHEP02(2014)110
  [arXiv:1310.1932 [hep-ph]].

\bibitem{stability-2}
  N.~Blinov and D.~E.~Morrissey,
  JHEP {\bf 1403}, 106 (2014)
  doi:10.1007/JHEP03(2014)106
  [arXiv:1310.4174 [hep-ph]].



\bibitem{suspect}
  A.~Djouadi, J.~L.~Kneur and G.~Moultaka,
  Comput.\ Phys.\ Commun.\  {\bf 176}, 426 (2007)
  doi:10.1016/j.cpc.2006.11.009
  [hep-ph/0211331].



\bibitem{higgsbounds}
  P.~Bechtle {\it et al.},
  Comput.\ Phys.\ Commun.\  {\bf 182}, 2605 (2011);
  Comput.\ Phys.\ Commun.\  {\bf 181}, 138 (2010).

\bibitem{higgssignals}
  P.~Bechtle {\it et al.},
  Eur.\ Phys.\ J.\ C {\bf 74}, 2711 (2014);
  P.~Bechtle {\it et al.},
  Comput.\ Phys.\ Commun.\  {\bf 181}, 138 (2010).

\bibitem{micromega}
  G.~Belanger {\it et al.},
  Comput.\ Phys.\ Commun.\  {\bf 182}, 842 (2011).


\bibitem{planck}
  P.~A.~R.~Ade {\it et al.}  [Planck Collaboration],
  arXiv:1303.5076 [astro-ph.CO].

\bibitem{cms-invisible}
  V.~Khachatryan {\it et al.} [CMS Collaboration],
  JHEP {\bf 1702}, 135 (2017)
  doi:10.1007/JHEP02(2017)135
  [arXiv:1610.09218 [hep-ex]].

\bibitem{lux_sd}
  D. S. Akerib {\it et al.} (LUX Collaboration), Phys. Rev. Lett. {\bf 118}, 021303 (2017) [arXiv:1608.07648 [astro-ph.CO]]

\bibitem{lux_si}
  D. S. Akerib {\it et al.} (LUX Collaboration), Phys. Rev. Lett. {\bf 116}, 161302 (2016) [arXiv:1602.03489 [hep-ex]]

\bibitem{pico2l}
  C.~Amole {\it et al.} [PICO Collaboration],
  Phys.\ Rev.\ D {\bf 93}, no. 6, 061101 (2016)
  doi:10.1103/PhysRevD.93.061101
  [arXiv:1601.03729 [astro-ph.CO]].

\bibitem{pico60}
  C.~Amole {\it et al.} [PICO Collaboration],
  Phys.\ Rev.\ D {\bf 93}, no. 5, 052014 (2016)
  doi:10.1103/PhysRevD.93.052014
  [arXiv:1510.07754 [hep-ex]].

\bibitem{icecube}
  M.~G.~Aartsen {\it et al.} [IceCube Collaboration],
  JCAP {\bf 1604}, no. 04, 022 (2016)
  doi:10.1088/1475-7516/2016/04/022
  [arXiv:1601.00653 [hep-ph]].


\bibitem{pxenon1t}
  E.~Aprile {\it et al.} [XENON Collaboration],
  JCAP {\bf 1604}, no. 04, 027 (2016)
  doi:10.1088/1475-7516/2016/04/027
  [arXiv:1512.07501 [physics.ins-det]].


\bibitem{checkmate}
  M.~Drees {\it et al.},
  Comput.\ Phys.\ Commun.\  {\bf 187}, 227 (2014).
  J.~S.~Kim {\it et al.},
  arXiv:1503.01123 [hep-ph].
  D.~Dercks, N.~Desai, J.~S.~Kim, K.~Rolbiecki, J.~Tattersall and T.~Weber,
  arXiv:1611.09856 [hep-ph].

\bibitem{mad5}
  J.~Alwall {\it et al.},
  JHEP {\bf 1407}, 079 (2014).

\bibitem{pythia}
  T.~Sj\'ostrand {\it et al.},
  Comput.\ Phys.\ Commun.\  {\bf 191}, 159 (2015)
  doi:10.1016/j.cpc.2015.01.024
  [arXiv:1410.3012 [hep-ph]].

\bibitem{delphes}
  J.~de Favereau,  {\it et al.},
arXiv:1307.6346 [hep-ex].

\bibitem{fastjet}
  M.~Cacciari, G.~P.~Salam and G.~Soyez,
  Eur.\ Phys.\ J.\ C {\bf 72}, 1896 (2012)
  [arXiv:1111.6097 [hep-ph]].

\bibitem{anti-kt}
  M.~Cacciari, G.~P.~Salam and G.~Soyez,
  JHEP {\bf 0804}, 063 (2008).

\bibitem{prospino}
  W.~Beenakker, M.~Klasen, M.~Kramer, T.~Plehn, M.~Spira and P.~M.~Zerwas,
  Phys.\ Rev.\ Lett.\  {\bf 83}, 3780 (1999)
  Erratum: [Phys.\ Rev.\ Lett.\  {\bf 100}, 029901 (2008)]
  doi:10.1103/PhysRevLett.100.029901, 10.1103/PhysRevLett.83.3780
  [hep-ph/9906298].

\bibitem{atlas-1402-7029}
  G.~Aad {\it et al.} [ATLAS Collaboration],
  JHEP {\bf 1404}, 169 (2014)
  doi:10.1007/JHEP04(2014)169
  [arXiv:1402.7029 [hep-ex]].


\bibitem{atlas-1501-07110}
  G.~Aad {\it et al.} [ATLAS Collaboration],
  Eur.\ Phys.\ J.\ C {\bf 75}, no. 5, 208 (2015)
  doi:10.1140/epjc/s10052-015-3408-7
  [arXiv:1501.07110 [hep-ex]].


\bibitem{atlas-conf-2013-049}
[ATLAS Collaboration], ATLAS-CONF-2013-049.

\bibitem{atlas-phys-2014-010-hl-3l}
[ATLAS Collaboration], ATL-PHYS-PUB-2014-010.


\end{thebibliography}
\end{document}